\documentclass[pra,twocolumn]{revtex4-2}

\usepackage[hidelinks]{hyperref}
\usepackage{amssymb}
\usepackage{array}
\usepackage{graphicx}
\usepackage{dsfont}
\usepackage{xcolor}

\begin{document}

\title{Evidence of a liquid phase in interacting Bosons at intermediate densities}

\author{Ian Jauslin}
\affiliation{\it Department of Mathematics, Rutgers University}
\email{ian.jauslin@rutgers.edu}

\begin{abstract}
In this paper, we present evidence for a liquid-like phase in systems of many interacting Bosons at intermediate densities.
The interacting Bose gas has been studied extensively in the low and high density regimes, in which interactions do not play a physically significant role, and the system behaves similarly to the ideal quantum gas.
Instead, we will turn our attention to the intermediate density regime, and report evidence that the system enters a strongly correlated phase where its behavior is markedly different from that of the ideal quantum gas.
To do so, we use the Simplified approach to the Bose gas, which was introduced by Lieb in 1963 and recently found to provide very accurate predictions for many-Boson systems at all densities.
Using this tool, we will compute predictions for the radial distribution function, structure factor, condensate fraction and momentum distribution, and show that they are consistent with liquid-type behavior.
\end{abstract}

\maketitle

\section{Introduction}

\indent
Since the early days of quantum mechanics, the Bose gas has been the subject of much interest, both from the theoretical\-~\cite{Bo24,Ei24,Le29,Bo47,Dy57,LHY57,Ef70,LY98,Ta08,BDZ08,YY09,Su11,CS16,NE17,FS20,BCS21,FS22} and the experimental\-~\cite{Ka38,AM38,AEe95,DMe95,KMe06,MMe07,BDZ08,CGe10,CS16,NE17,FAe21} communities.
Despite its relative simplicity, it exhibits a rich phenomenology: it forms a Bose-Einstein condensate at low temperatures\-~\cite{Bo24,Ei24,AEe95,DMe95}, and, with the advent of cold-atom physics\-~\cite{PM82,CHe85,AAe88,BML00} and the possibility of studying Bose gasses in the lab with ever increasing precision, there have been many successes in probing its phase diagram and understanding its exotic quantum phase transitions\-~\cite{AEe95,DMe95,MMe07,SPP07,BDZ08}.

\indent
Whereas much attention has been payed to the behavior of Bose gasses at very low and high densities, where the system behaves similarly to the ideal quantum gas, in this paper, we shall turn our attention to the intermediate density regime, for which we have found evidence of behavior that differs significantly from the ideal quantum gas, and bears resemblance to a liquid-type phase.
Until recently, theoretical tools, such as Bogolyubov theory\-~\cite{Bo47,LY98,ZB01,LSe05,YY09,FS22} or renormalization group techniques\-~\cite{Be95,CG14,BFe17}, developed to understand the behavior of the Bose gas have been based on perturbing non-interacting systems.
As such, these methods are ill-suited to understanding the strongly coupled behavior emerging in the intermediate density regime.

\indent
Instead, we will use the ``Simplified Approach to the Bose gas'', which was introduced in a paper by Lieb from 1963\-~\cite{Li63,LS64,LL64}, and was recently found to yield very accurate predictions at all densities\-~\cite{CJL20,CJL21,CHe21,Ja22}.
This has allowed us to probe the behavior of Bose gasses in a range of densities that had, until now, only been accessible to Quantum Monte-Carlo simulations.
In doing so, we have found numerical evidence for a liquid-like phase in a range of densities that is large enough for the interactions to become important, but not so large as to break into the mean-field regime.
This is, as far as we know, a new prediction, which shows that there is non-trivial behavior in interacting Bose gasses at intermediate densities, and may be investigated experimentally.

\indent
More specifically, we have studied predictions for the radial distribution function (i.e. the spherical average of the two-point correlation function), the structure factor (i.e. the Fourier transform of the radial distribution function), the condensate fraction, and the momentum distribution (i.e. the average number of particles in the state $e^{i\mathbf k\mathbf x}$).
We have found that the radial distribution function is monotone increasing for small densities, and that, beyond a first critical density $\rho_*$, a local maximum emerges, see Figures\-~\ref{fig:2pt} and\-~\ref{fig:2pt_max}.
There is thus a length scale at which it is more likely to find pairs of particles, which is consistent with liquid behavior.
Conversely, the structure factor is monotone at very high densities, and, lowering the density, we find that for densities smaller than a second critical density $\rho_{**}>\rho_*$, it develops a local maximum, see Figures\-~\ref{fig:2pt_fourier} and\-~\ref{fig:2pt_fourier_max}.
These critical densities also appear rather close to inflection points of the condensate fraction as a function of density, see Figure\-~\ref{fig:condensate}.
We have also investigated the momentum distribution, and found that it increases sharply near $\rho_*$, see Figure\-~\ref{fig:Nk}.
This is clear evidence for non-trivial behavior in the range of densities $\rho_*<\rho<\rho_{**}$, which shares some similarities to classical liquids\-~\cite{HM88}.

\indent
These results complete the phase diagram of the Bose gas.
At low densities, the interactions between particles are weak, and the system behaves similarly to the ideal quantum gas\-~\cite{Bo47,Dy57,LHY57,LY98,YY09,FS20,FS22,LSY00,LS02,NRS16,BBe18,BBe19,DSY19,BBe20,DS20,BSS22,NNe22,Sc22}.
At high densities, the particles are so close that the effect of neighboring particles is approximately a uniform background field: this is a mean-field phase\-~\cite{LSe05,Se11,PPS20,Bo22}, and behaves formally as an ideal quantum gas in a field.
The results in this paper show that, in between these two regimes, there is evidence for a new kind behavior.
It is worth pointing out that, in the case of a gas with hard-core repulsion, the mean-field regime does not exist, and the intermediate density regime considered here corresponds to the high density phase of the hard-core Bose gas.
\bigskip

\indent
The model we will consider throughout this paper is a systems of many Bosons interacting via a spherically symmetric, repulsive, pair potential, whose Hamiltonian is
\begin{equation}
  H=-\frac12\sum_{i=1}^N\Delta_i+\sum_{1\leqslant i<j\leqslant N}v(|x_i-x_j|)
\end{equation}
which we will consider in the thermodynamic limit $N,V\to\infty$ with $N/V=\rho$ fixed ($V$ is the volume).
The Simplified approach consists in reducing the computation of thermodynamic observables of this system to solving a non-linear, non-local effective equation\-~(\ref{bigeq}) on $\mathbb R^3$, by making an (as of yet uncontrolled) approximation, see\-~\cite{CJL20,CJL21,CHe21,Ja22} for more details.
Doing so comes at a cost, and there are several important limitations to the method.
In particular, the Simplified approach seems only to be useful to compute the ground state of Bose gasses, which means that we can probe the extremely low-temperature regime of the phase diagram, but not higher temperatures.
In addition, the high-density predictions of the Simplified approach have been shown\-~\cite{CJL20,CJL21,CHe21} to be accurate only in the case of purely repulsive interactions of positive type, that is, to potentials $v$ that are $\geqslant 0$ and whose Fourier transform is also $\geqslant 0$.
Such potentials are not rare: given any non-negative function $f$, the potential $v(x)=f\ast f(x)\equiv \int dy\ f(x-y)f(y)$ satisfies the two requirements.
Finally, we will assume that the interaction is spherically symmetric, as that greatly simplifies the numerical solution of the effective equation.
Under these restrictions, the Simplified approach has been found to be extremely accurate\-~\cite{CHe21} when compared to analytical predictions and to Quantum Monte Carlo simulations.

\indent
The Simplified approach actually provides a family of equations with varying levels of approximation.
In most of this paper we will use the ``Big equation''\-~\cite{CHe21}, which provides the best compromise between computational efficiency and accuracy.
It is defined as
\begin{equation}
  \begin{array}{r@{\ }l}
    -\Delta u(\mathbf x)
    =&
    (1-u(\mathbf x))
    \big(v(\mathbf x)-2\rho u\ast S(\mathbf x)
    +\\[0.1cm]&+
    \rho^2 u\ast u\ast S(\mathbf x)-2u\ast(u(u\ast S))(\mathbf x)
    \big)
  \end{array}
  \label{bigeq}
\end{equation}
in which $\ast$ is the convolution operator, $\rho$ is the density, $v$ is the potential, and
\begin{equation}
  S(\mathbf x):=(1-u(\mathbf x))v(\mathbf x)
  .
  \label{S}
\end{equation}
The unknown $u$ is related to the two-point of correlation function of the ground-state wavefunction $\psi_0$, viewed as a probability distribution:
\begin{equation}
  u(\mathbf x_1-\mathbf x_2)=1-\lim_{N,V\to\infty}\frac{V^2\int d\mathbf x_3\cdots d\mathbf x_N\ \psi_0(\mathbf x_1,\cdots,\mathbf x_N)}{\int d\mathbf y_1\cdots d\mathbf y_N\ \psi_0(\mathbf y_1,\cdots,\mathbf y_N)}
\end{equation}
in terms of which we can compute the ground state energy per particle:
\begin{equation}
  e=\frac\rho2\int d\mathbf x\ (1-u(\mathbf x))v(\mathbf x)
  .
  \label{energy}
\end{equation}

\indent
The computation of the momentum distribution will actually be done in a different approximation, as the Big equation leads to significant numerical difficulties for that observable.
Instead, we will consider another of the equations of the Simplified approach: the ``Medium equation''\-~\cite{CHe21}, which is less accurate, but much easier to solve numerically.
It is obtained from the Big equation by neglecting the $2u\ast(u(u\ast S))$ term, and dropping the $u(\mathbf x)$ in the $(1-u(\mathbf x))$ prefactor except in front of $v$:
\begin{equation}
  -\Delta u(\mathbf x)
  =
  (1-u(\mathbf x))v(\mathbf x)
  -2\rho u\ast S(\mathbf x)
  +\rho^2 u\ast u\ast S(\mathbf x)
  .
  \label{medeq}
\end{equation}
We will also use the less accurate Medium equation as a check on the predictions of the Big equation: qualitative phenomena that are visible in both approaches have a good chance of holding for the exact, unapproximated many-body Bose gas as well.
Conversely, when the quantitative predictions disagree, we will take that as an indication that the quantitative predictions are not to be taken too seriously.

\indent
Throughout this paper, we will use the interaction potential
\begin{equation}
  v(\mathbf x)=8e^{-|\mathbf x|}
\end{equation}
which is of positive type (its Fourier transform is non-negative).
There is no particular reason why this potential is used rather than another spherically symmetric, positive type function, and it is chosen in this way merely for the sake of definiteness.
\bigskip

\indent
The rest of the paper is structured as follows.
In Section\-~\ref{sec:results}, we present the main results, and discuss the prediction of the Simplified approach for the radial distribution function, structure factor, condensate fraction and momentum distribution, and find that these consistently show non-trivial behavior for intermediate densities, which is consistent with a liquid-type phase.
In Appendix\-~\ref{app:medeq}, we present the corresponding predictions for the Medium equation.
In Appendix\-~\ref{app:simplesolv}, we discuss the numerical computation of the solution of the Big and Medium equations, which were carried out using the {\tt simplesolv}\-~\cite{ss} tool developed for this purpose, and released under a free software license.

\section{Numerical analysis of the intermediate density phase}\label{sec:results}
\subsection{Radial distribution}
\indent
We define the radial distribution function as the spherical average of the normalized two-point correlation function:
\begin{equation}
  g(r):=\frac1{\rho^24\pi r^2}\int d\mathbf y\ \delta(|\mathbf y|-r)\sum_{i,j=1}^N\left<\delta(\mathbf y-\mathbf x_i)\delta(\mathbf x_j)\right>
  .
  \label{C2}
\end{equation}
Normalized in this way, $g\to1$ as $r\to\infty$.
To compute $g$, we use the fact that, denoting the energy of the system by $E_0$,
\begin{equation}
  \frac12\sum_{i,j=1}^N\left<\delta(\mathbf y-\mathbf x_i)\delta(\mathbf x_j)\right>=\frac{\delta E_0}{\delta v(\mathbf y)}
\end{equation}
and use the prediction of the Big equation for the energy of the Bose gas to compute $E_0$.
\bigskip

\begin{figure}
  \hfil\includegraphics[width=\columnwidth]{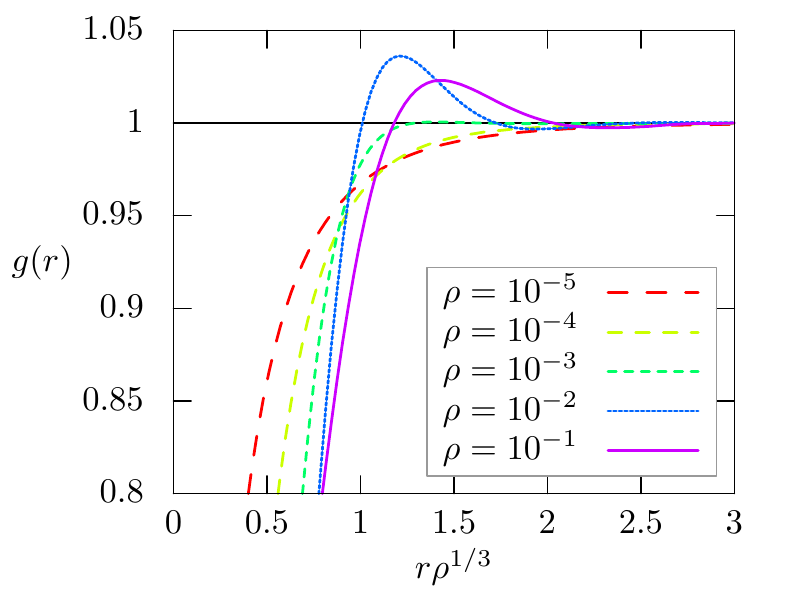}

  \caption{
    The prediction of the Big equation for the radial distribution function as a function of $r\rho^{1/3}$, for various values of $\rho$.
    Note that $\rho^{-1/3}$ is the length scale of the average inter-particle distance.
    As $\rho$ increases the radial distribution function develops a peak above $1$ that is not present for smaller densities.
    As the density is increased further, the height of the peak goes down.
    A similar plot for the Medium equation is in Figure\-~\ref{fig:2pt_medeq}.
  }
  \label{fig:2pt}
\end{figure}

\indent
The prediction for the radial distribution function for the Big equation is shown in Figure\-~\ref{fig:2pt}.
At low densities, the maximum of $g$ is $1$, that is, it is attained as $r\to\infty$.
As the density is increased, there is a transition to a regime in which the maximum is greater than $1$, and is attained at a finite value $r_*$.
In such cases, the length scale $r_*$ is a preferred inter-particle spacing, which shows that there is short-range order in the system.
This maximum quickly dissipates as $r$ increases, thus showing that there is no long-range order, which is consistent with the behavior of a liquid phase.
\bigskip

\begin{figure}
  \hfil\includegraphics[width=\columnwidth]{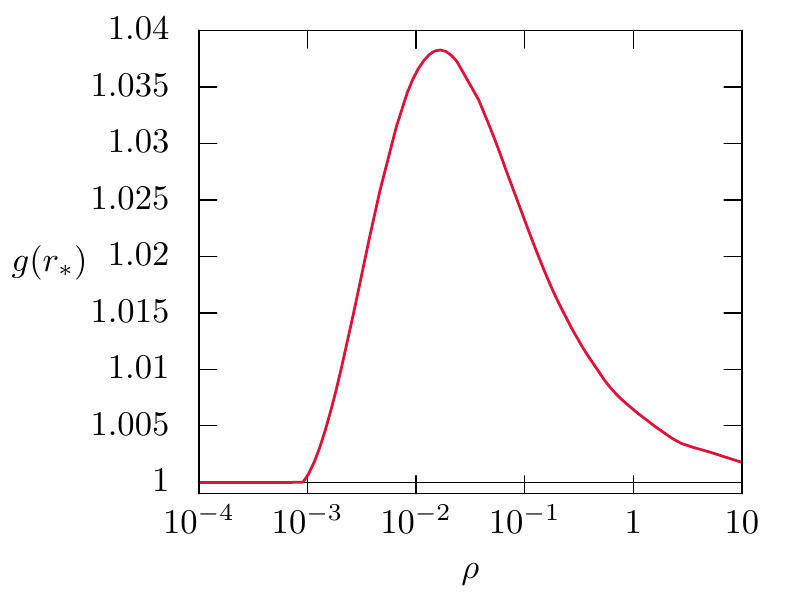}

  \caption{
    The prediction of the Big and Medium equations for the maximum of the radial distribution function as a function of $\rho$.
    There is a clear transition from a low density regime in which the maximum is $1$ to a high density regime in which the maximum is $>1$.
    The critical density at which the transition occurs is approximately $\rho_*=0.9\times 10^{-3}$.
  }
  \label{fig:2pt_max}
\end{figure}

\indent
The transition is even clearer in Figure\-~\ref{fig:2pt_max}, which shows the prediction of the maximum of the radial distribution function as a function of the density.
We see a clear transition from a low density regime in which the maximum $g(r_*)$ of $g$ is 1 to a high-density regime in which $g(r_*)>1$.
This occurs at a density $\rho_*\approx 0.9\times10^{-3}$, though the precise value of $\rho_*$ should not be taken too seriously.
Indeed, as is seen in Figure\-~\ref{fig:2pt_max_medeq} in Appendix\-~\ref{app:medeq}, the qualitative behavior of the Medium equation is similar to that of the Big equation, but the value of $\rho_*$ is off by a factor of $\approx2$.
Since the Big and Medium equation are two different levels of approximation of the many-body Bose gas, this is evidence that the Bose gas has a transition from $g(r_*)=1$ to $g(r_*)>1$, through the precise value of $\rho_*$ may differ from that of the Big equation.

\subsection{Structure factor}

\indent
The structure factor is defined in terms of the Fourier transform of the radial distribution function $g$\-~\cite{HM88}:
\begin{equation}
  \mathcal S(|\mathbf k|):=1+\rho\int d\mathbf x\ e^{i\mathbf k\mathbf x}(g(|\mathbf x|)-1)
  .
\end{equation}
The structure factor is of interest as it is directly observable in scattering experiments\-~\cite{HM88}.
\bigskip

\begin{figure}
  \hfil\includegraphics[width=\columnwidth]{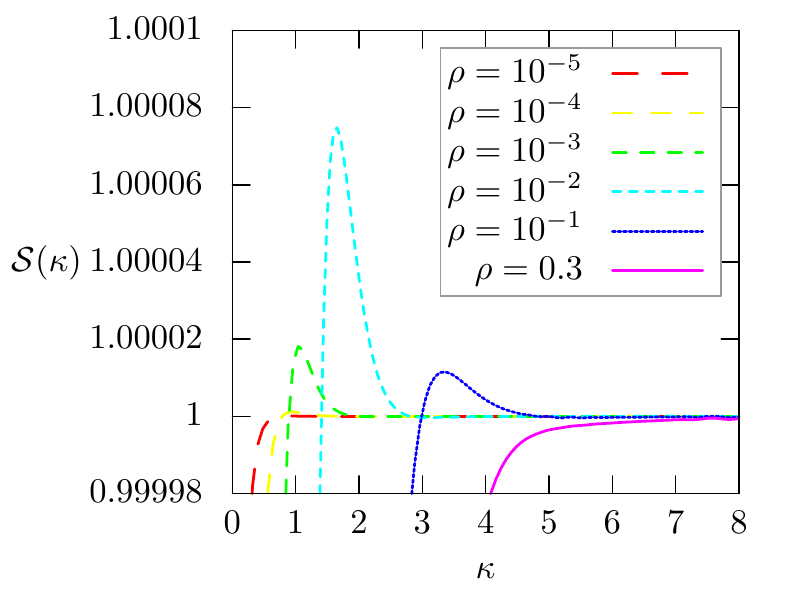}

  \caption{
    The prediction of the Big equation for the structure factor as a function of $\kappa\equiv|\mathbf k|$, for a wide range of values of $\rho$.
    At small densities, the structure factor has a maximum that is just slightly above one (not visible in the figure), and as the density increases, this maximum becomes more and more pronounced, and then decreases.
    Above a certain density, the maximum disappears entirely.
    A similar plot for the Medium equation is in Figure\-~\ref{fig:2pt_fourier_medeq}.
  }
  \label{fig:2pt_fourier}
\end{figure}

\indent
The prediction for the structure factor for the Big equation is shown in Figure\-~\ref{fig:2pt_fourier}.
We find that, as the density increases, the maximum of the structure factor increases, and its standard deviation becomes smaller.
This bump is far from being a Bragg peak, as there is no long range order, nevertheless, the sharpening of the maximum indicates increased correlations\-~\cite{To18}, which is consistent with liquid-type behavior.
As the density is increased further, this bump disappears, as the system transitions to a high-density mean-field regime.
\bigskip

\begin{figure}
  \hfil\includegraphics[width=\columnwidth]{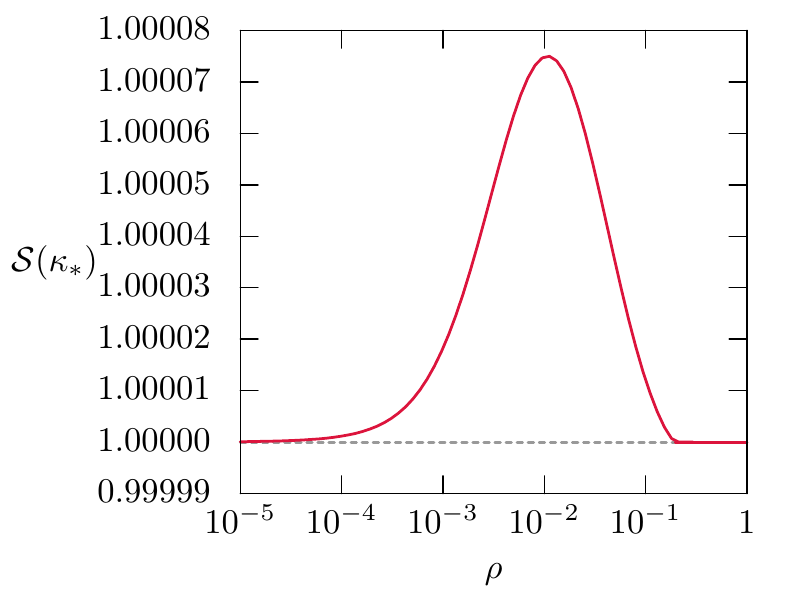}

  \caption{
    The prediction of the Big equation for the maximum of the structure factor as a function of $\rho$.
    As the density increases, the maximum of $\mathcal S$ first increases, then reaches a maximum, and decreases anew.
    Beyond a density $\rho_{**}\approx 0.2$, the maximum of $\mathcal S$ is equal to 1.
  }
  \label{fig:2pt_fourier_max}
\end{figure}

\indent
In Figure\-~\ref{fig:2pt_fourier_max}, we plot the maximum of $\mathcal S$ as a function of $\rho$, where we see that the maximum increases smoothly until it reaches a maximum, and then decreases anew.
Beyond a second critical density, $\rho_{**}\approx0.2$, the local maximum disappears, and the maximum of $\mathcal S$ is pushed off to $\infty$.
Again, the value of this critical density should not be taken too seriously, as is indicated by a comparison with the prediction of the Medium equation, see Figure\-~\ref{fig:2pt_fourier_max_medeq}.

\subsection{Condensate fraction}

\indent
The condensate fraction is the proportion of particles in the Bose-Einstein condensate:
\begin{equation}
  \eta=
  \frac1N\sum_{i=1}^N\left<P_0^{(i)}\right>
\end{equation}
where $P_0^{(i)}$ is the projector onto the subspace in which the $i$-th particle is in the constant state $\frac1{\sqrt V}$.
To compute it, we use the Feynman-Hellman theorem and express $\eta$ as a derivative of the ground state energy of an effective Hamiltonian, which we compute using the Big equation\-~\cite{CJL21,CHe21}.
\bigskip

\begin{figure}
  \hfil\includegraphics[width=\columnwidth]{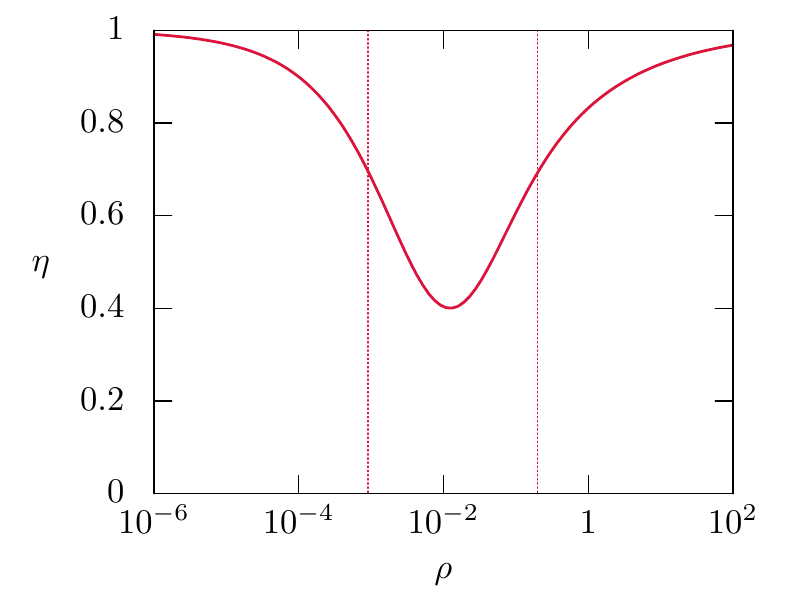}

  \caption{
    The prediction of the Big equation for the condensate fraction as a function of $\rho$.
    As the density increases, the condensate fraction first decreases, then reaches a minimum, and increases anew.
    The dotted vertical lines correspond to the critical densities $\rho_*=0.9\times10^{-3}$ and $\rho_{**}=0.2$.
    The curve has inflection points that are somewhat near $\rho_*$ and $\rho_{**}$.
  }
  \label{fig:condensate}
\end{figure}

\indent
We plot the condensate fraction as a function of the density in Figure\-~\ref{fig:condensate}.
As $\rho\to0$, $\eta\to1$, that is, there is complete Bose-Einstein condensation at zero density.
As the density is increased, $\eta$ decreases, then reaches a minimum, and then increases back towards $1$.
There are two inflection points, which occur somewhat close to the critical densities $\rho_*\approx0.9\times10^{-3}$ and $\rho_{**}\approx0.2$.

\subsection{Momentum distribution}

\indent
The condensate fraction is defined using the projector onto the constant state, which is the ground state of the non-interacting system (the Laplacian).
The momentum distribution is determined from the occupation number of the {\it excited} states of the Laplacian, namely $e^{i\mathbf k\mathbf x}$ (note that this is different from studying the excitation spectrum of the Bose gas; our computation is restricted to the ground state).
Specifically, we define the number of particles with momentum $|\mathbf k|\equiv\kappa$ as
\begin{equation}
  N_{\kappa}:=\int d\mathbf k\ \delta(|\mathbf k|-\kappa)\sum_{i=1}^N\left<P_{\mathbf k}^{(i)}\right>
\end{equation}
where $P_{\mathbf k}^{(i)}$ is the projector onto the subspace in which the $i$-th particle is in the state $\frac1{\sqrt V}e^{i\mathbf k\mathbf x}$.
Thus, $N_{\kappa}$ is the integral over the sphere of radius $\kappa$ of the number of particles in the state $e^{i\mathbf k\mathbf x}$.
In particular, $\eta=N_0/N$.
(The momentum distribution is then defined as $\mathcal M(\kappa):=N_\kappa/(4\pi\kappa^2\rho)$, but, in the following, we shall show results for $N_\kappa$ instead.)
\bigskip

\indent
As is explained in more detail in Appendix\-~\ref{app:simplesolv}, the numerical solution of the Big equation is less accurate than that of the Medium equation, and the computation of the momentum distribution for the Big equation leads to large numerical artifacts.
We will therefore focus on the Medium equation.
We will compare the prediction of the Medium equation to that of Bogolyubov theory\-~\cite[Appendix\-~A]{LSe05}:
\begin{equation}
  N_{\kappa}^{(\mathrm{Bog})}=\frac12\left(\frac{\kappa^2+8\pi\rho a}{\kappa^2(\kappa^2+16\pi\rho a)}-1\right)
\end{equation}
where $a$ is the scattering length of the potential.

\begin{figure}
  \hfil\includegraphics[width=\columnwidth]{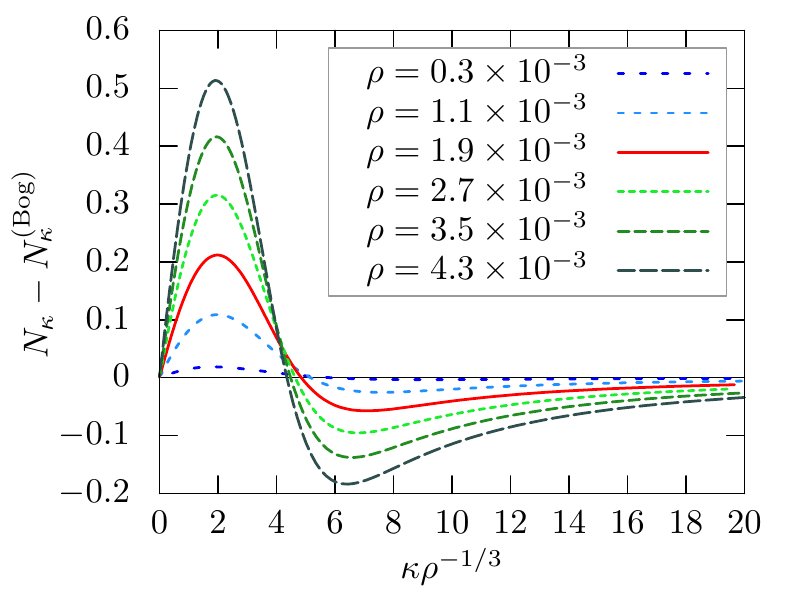}

  \caption{
    The difference between the predictions of the Medium equation and Bogolyubov theory for the spherical integral of the occupation number in the state $e^{i\mathbf k\mathbf x}$ as a function of $\kappa\equiv|\mathbf k|$ for densities near the critical density $\bar\rho_*=1.9\times10^{-3}$.
    The solid line corresponds to $\rho=\bar\rho_*$.
    As the density approaches $\bar\rho_*$, the difference in the predictions grows quickly.
    We also find that Bogolyubov underestimates the occupation number for small $\kappa$ and overestimates it for large $\kappa$.
  }
  \label{fig:Nk}
\end{figure}

\indent
We plot the difference between the prediction for $N_\kappa$ of the Medium equation and of Bogolyubov theory in Figure\-~\ref{fig:Nk}.
We find that this difference increases sharply near the critical density $\bar\rho_*$ (for the Medium equation, the transition density is $\bar\rho_*\approx1.9\times10^{-3}$).
In addition, we find that Bogolyubov theory underestimates $N_\kappa$ for small $\kappa$ and overestimates it for larger $\kappa$.

\section{Conclusion}
\indent
We have shown evidence for the existence of a non-trivial phase in interacting Bose gasses in a range of densities that are neither very small nor very large.
More specifically, we have shown that there exist two critical densities, $\rho_*<\rho_{**}$ such that, for $\rho_*<\rho<\rho_{**}$, both the radial distribution function and the structure factor have a maximum, see Figures\-~\ref{fig:2pt}-\ref{fig:2pt_fourier_max}.
Outside this range of densities, either the radial distribution function or the structure factor does not have a maximum.
This suggests a behavior that is similar to that of a classical liquid\-~\cite{HM88} for $\rho_*<\rho<\rho_{**}$.
In addition, these critical densities are near inflection points of the condensate fraction, see Figure\-~\ref{fig:condensate}.
Even though the evidence is insufficient to confidently claim that the system is in a liquid phase, it seems clear that there is non-trivial behavior in this intermediate range of densities.

\indent
To study this range of densities, we have used the Simplified approach, which is a method to study the ground state of repulsive Bose gasses with positive-type pair potentials.
It would be interesting to check these predictions using Quantum Monte-Carlo simulations (as was done for the radial distribution function in\-~\cite{CHe21}), and perhaps even in experiments.
This paper shows clear evidence that the behavior in the intermediate density regime may be worth investigating further, both theoretically and experimentally.

\begin{acknowledgements}
The author thanks Elliott H. Lieb, Eric A. Carlen and Markus Holzmann for many valuable discussions.
The author acknowledges support from the Simons Foundation, Grant Number 825876.
\end{acknowledgements}

\appendix

\section{Predictions of the Medium equation}\label{app:medeq}
\begin{figure}
  \hfil\includegraphics[width=\columnwidth]{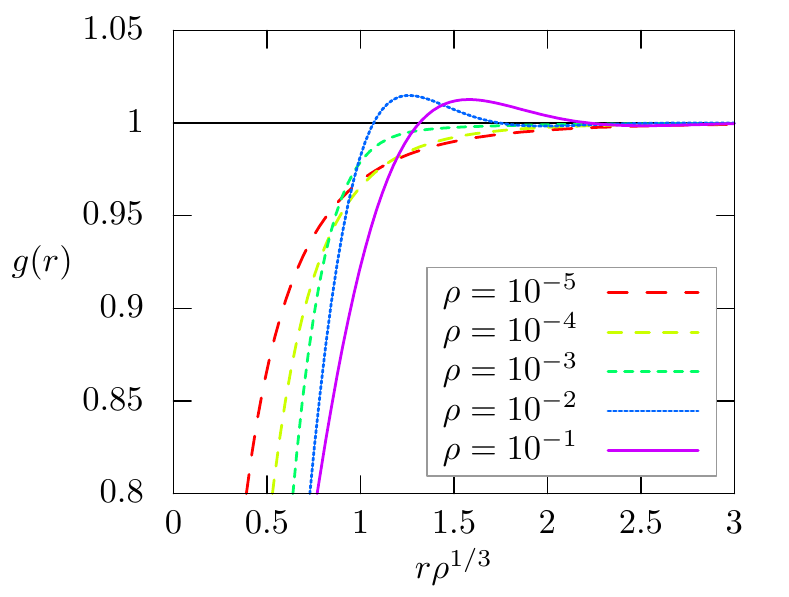}

  \caption{
    The prediction of the Medium equation for the radial distribution function as a function of $r\rho^{1/3}$, for various values of $\rho$.
    The Medium equation reproduces the qualitative behavior of the Big equation in Figure \ref{fig:2pt}.
  }
  \label{fig:2pt_medeq}
\end{figure}

\begin{figure}
  \hfil\includegraphics[width=\columnwidth]{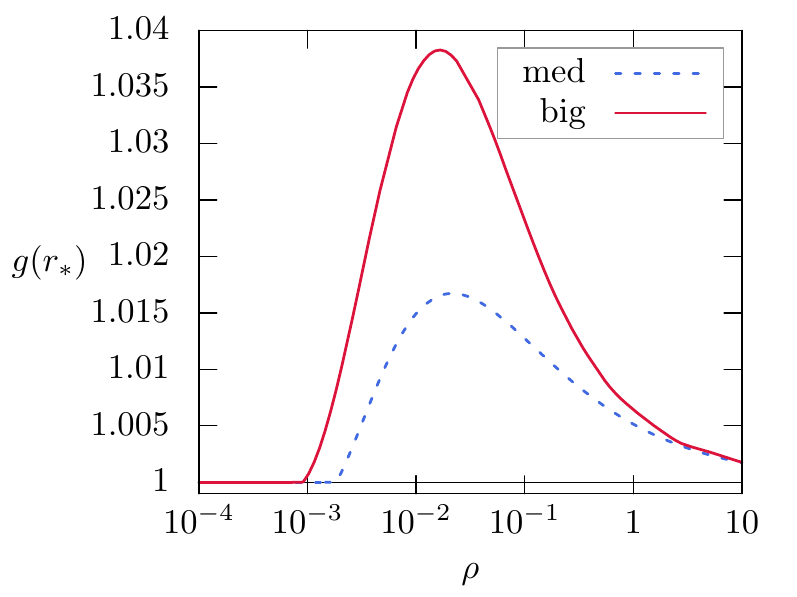}

  \caption{
    The prediction of the Big and Medium equations for the maximum of the radial distribution function as a function of $\rho$.
    The Medium equation is qualitatively similar to the Big equation, but the location of the transition as well as the height of the maximum differ significantly.
    The critical density at which the transition occurs for the Medium equation is $\bar\rho_*\approx1.9\times 10^{-3}$.
  }
  \label{fig:2pt_max_medeq}
\end{figure}

\begin{figure}
  \hfil\includegraphics[width=\columnwidth]{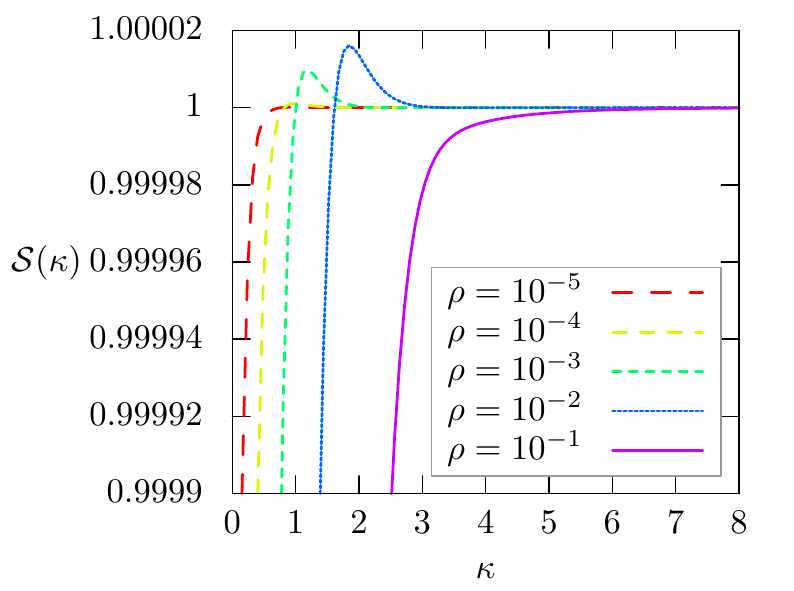}

  \caption{
    The prediction of the Medium equation for the structure factor as a function of $\kappa\equiv|\mathbf k|$, for a wide range of values of $\rho$.
    The Medium equation reproduces the qualitative behavior of the Big equation in Figure \ref{fig:2pt_fourier}.
  }
  \label{fig:2pt_fourier_medeq}
\end{figure}

\begin{figure}
  \hfil\includegraphics[width=\columnwidth]{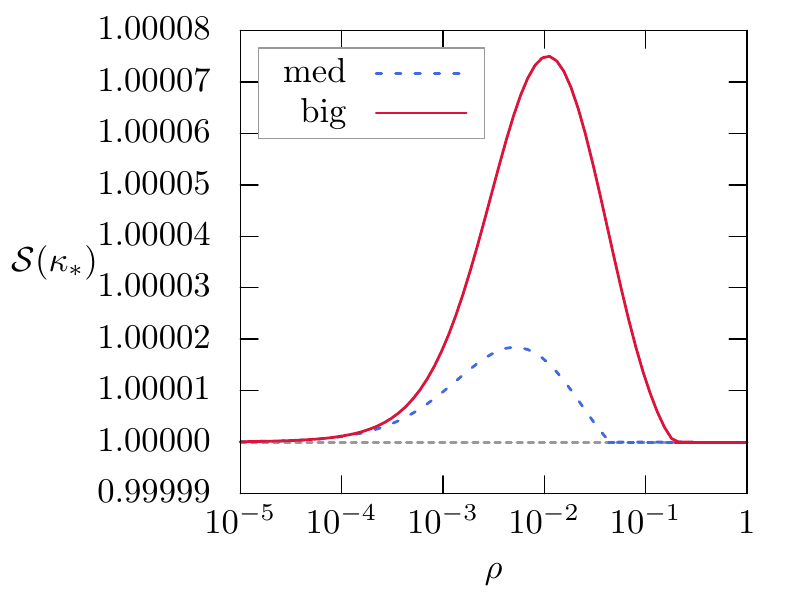}

  \caption{
    The prediction of the Big and Medium equations for the maximum of the structure factor as a function of $\rho$.
    The Medium equation is qualitatively similar to the Big equation, but the location of the transition as well as the height of the maximum differ significantly.
    The critical density at which the transition occurs for the Medium equation is $\bar\rho_{**}\approx0.05$.
  }
  \label{fig:2pt_fourier_max_medeq}
\end{figure}

\begin{figure}
  \hfil\includegraphics[width=\columnwidth]{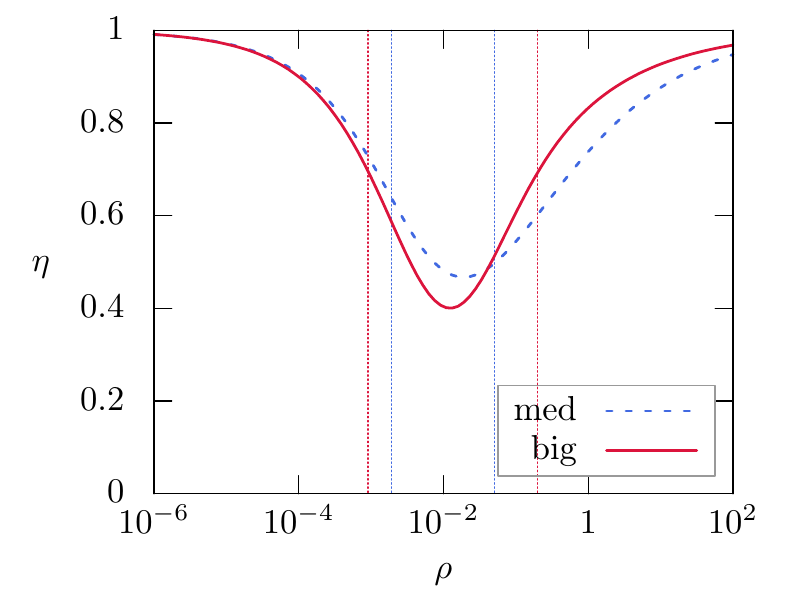}

  \caption{
    The prediction of the Big and Medium equations for the condensate fraction as a function of $\rho$.
    The dotted vertical lines correspond to the critical densities $\rho_*=0.9\times10^{-3}$ and $\rho_{**}=0.2$ for the Big equation and $\bar\rho_*=1.9\times10^{-3}$ and $\bar\rho_{**}=0.05$ for the Medium equation.
    For the Big equation, the curve has inflection points at $\rho_*$ and $\rho_{**}$.
    For the Medium equation, they are off.
  }
  \label{fig:condensate_medeq}
\end{figure}

In this appendix, we show plots of the predictions of the Medium equation for the results discussed above for the Big equation, see Figures\-~\ref{fig:2pt_medeq}-\ref{fig:condensate_medeq}.

\section{{\tt simplesolv}: a tool to solve the equations of the Simplified approach}\label{app:simplesolv}
\indent
To compute the numerical solution to the equations of the Simplified approach (such as the Big and Medium equations) we developed a tool called {\tt simplesolv}\-~\cite{ss}, written using the {\it Julia} programming language\-~\cite{Julia}, and released under the Apache 2.0 license, a free software license that allows free use, distribution, and modifications.
It is designed to compute the solution of any of the equations of the Simplified approach as well as a variety of observables, such as the energy, the condensate fraction, the two-point correlation function and its Fourier transform, the momentum distribution, and the compressibility.
\bigskip

\indent
In this appendix, we sketch the algorithm used to carry out the computation.
A more detailed explanation is available in the documentation bundled with the {\tt simplesolv} package\-~\cite{ss}.
\bigskip

\indent
The only observable that is directly accessible from the solution of the Big or Medium equations is the ground state energy per particle\-~(\ref{energy}).
To compute all other observables, we use the Feynman-Hellman theorem to reduce the computation to that of the energy of an auxiliary Hamiltonian, which leads to auxiliary Big and Medium equations.
We can thus reduce the computation of many observables to that of the energy.
\bigskip

\indent
We begin by describing the algorithm for the Medium equation, as it is simpler.
The Medium equation\-~(\ref{medeq}) can be rewritten as
\begin{equation}
  -\Delta u=S-2\rho u\ast S+\rho^2 u\ast u\ast S
  .
\end{equation}
In this form, it involves convolutions, but no products, so it has a simple expression in Fourier space:
\begin{equation}
  \mathbf k^2\hat u=\hat S-2\rho\hat S\hat u+\rho^2\hat S\hat u^2
  ,\quad
  \hat u(\mathbf k):=\int d\mathbf x\ e^{i\mathbf k\mathbf x}u(\mathbf x)
\end{equation}
with
\begin{equation}
  \hat S(\mathbf k):=\int d\mathbf x\ e^{i\mathbf k\mathbf x}S(\mathbf x)
  =\hat v(\mathbf k)+\frac1{8\pi^3}\hat v\ast\hat u(\mathbf k)
  .
\end{equation}
This equation thus only involves a single in $\hat S$.
To compute it numerically, we use a Gauss quadrature.
First of all, we assume radial symmetry and work in spherical coordinates, so the integral can be expressed in terms of an integral over $[0,\infty)$:
\begin{equation}
  \hat v\ast \hat u(|\mathbf k|)=\frac{2\pi}{|\mathbf k|}\int_0^\infty dt\ t\hat u(t)\int_{||\mathbf k|-t|}^{|\mathbf k|+t}ds\ s\hat v(s)
  .
\end{equation}
Next, we compactify the interval using the map $\kappa\mapsto1/(\kappa+1)$, which maps $[0,\infty)$ to $(0,1]$, and use a Gauss-Legendre quadrature in that interval.
The reason we compactify the interval, rather than use a quadrature defined directly on $[0,\infty)$, is that $\hat u$ decays algebraically (as $|\mathbf k|^{-2}$\-~\cite{CJL20}), which rules out using Gauss-Hermite and Gauss-Laguerre quadratures.
Proceeding in this way, we approximate
\begin{equation}
  \hat v\ast\hat u(\kappa_i)
  \approx
  \frac1{4\pi^3}\sum_{j=1}^N w_j\frac{(1-r_j)\hat u(\kappa_j)H(\kappa_i,\kappa_j)}{(1+r_j)^3}
\end{equation}
where $N$ is the {\it order} of the approximation, $(w_j,r_j)$ are the {\it weights} and {\it abscissa} of the Gauss-Legendre quadrature (which are universal and can be found in tables or standard software packages), and
\begin{equation}
  \kappa_i:=\frac{1-r_i}{1+r_i}
  ,\quad
  H(\kappa,t):=\frac{2\pi}\kappa\int_{|\kappa-t|}^{\kappa+t}ds\ s\hat v(s)
  .
\end{equation}
Having made this approximation, the Medium equation reduces to a system of equations for $\hat u(\kappa_i)$ for $i\in\{1,\cdots,N\}$, which we solve using the Newton algorithm.

\indent
Gauss quadratures can be proved to converge exponentially in $N$ for analytic functions\-~\cite{PTe08} so the algorithm converges exponentially in $N$ as long as $\hat u$ is analytic (algebraically if it is only $\mathcal C^p$).

\indent
For the plots in this paper, we have used $N=100$ or $N=200$.
\bigskip

\indent
The Big equation poses a more significant challenge.
Indeed, in Fourier space, (\ref{bigeq}) becomes
\begin{equation}
  \begin{array}{>\displaystyle l}
    -\mathbf k^2\hat u
    =
    \hat S
    -2\rho\hat S\hat u+\rho^2\hat S\hat u^2
    -\frac1{4\pi^3}\hat u(\hat u\ast(\hat S\hat u))
    -\\[0.1cm]\indent-
    \frac1{8\pi^3}\hat u\ast\left(
      -2\rho\hat S\hat u+\rho^2\hat S\hat u^2
      -\frac1{4\pi^3}\hat u(\hat u\ast(\hat S\hat u))
    \right)
    .
  \end{array}
\end{equation}
This involves many more convolutions in Fourier space than the Medium equation.
Whereas, for the Medium equation, using Gauss quadratures reduces the equation to a discrete system of equations, this is not the case for the Big equation.
Instead, we need an interpolation scheme to approximate the value of $\hat u$ in between the points $\kappa_i$.
To do so, we will use a Chebyshev polynomial expansion, but we must be careful in doing so alongside the compactification.
Indeed, we must take care to ensure that the polynomial goes to 0 at the edge of the compactified interval that corresponds to $\infty$, and that it does so at the appropriate rate.
To do so, instead of expanding $\hat u$, we expand $(1+|\mathbf k|)^2\hat u$, which does not decay at infinity.
In addition, because $\hat u$ is not necessarily approximated well by a polynomial uniformly over the entire range $[0,\infty)$, we split it up into intervals called {\it splines}, and perform the polynomial expansion in each spline independently.
(In addition to improving the precision, this gives us a simple check of the accuracy of the computation: neighboring splines must continue one another continuously, which allows us to spot numerical inaccuracies when this is not the case.)
Having approximated $\hat u$ by a polynomial, we compute integrals using Gauss-Legendre quadratures as before.
We denote the number of splines by $J$, the order of the Chebyshev polynomial expansion in each spline by $P$, and the order of the Gauss quadratures by $N$.

\indent
The Chebyshev polynomial expansion can be proved to converge exponentially in $P$ for analytic functions\-~\cite{PTe08}, so the algorithm converges exponentially in $N$ and in $P$ as long as $u$ is analytic.
However, it is computationally much heavier than the algorithm for the Medium equation, which restricts the values of $P,N$ and $J$ we can use in practice (all computations were run on a laptop computer).
Therefore, the numerical solution of the Big equation is more time-consuming, and, for some observables, less accurate than the solution of the Medium equation.

\indent
For the plots in this paper, we have used $J=10$, $P=8$ and $N=12$ or $J=15$, $P=12$ and $N=18$.

\bibliographystyle{apsrev4-2}
\bibliography{bibliography}

\end{document}